\documentclass[aps,prl,showpacs,twocolumn,superscriptaddress]{revtex4}
\usepackage{bm,color}
\usepackage{graphicx}
\usepackage{amsmath, amssymb}
\begin{document}
\title{Dynamical Switching of Magnetic Topology in Microwave-Driven Itinerant Magnet}
\author{Rintaro Eto}
\affiliation
{Department of Applied Physics, Waseda University, Okubo, Shinjuku-ku, Tokyo 169-8555, Japan}
\author{Masahito Mochizuki}
\affiliation
{Department of Applied Physics, Waseda University, Okubo, Shinjuku-ku, Tokyo 169-8555, Japan}
\begin{abstract}
We theoretically demonstrate microwave-induced dynamical switching of magnetic topology in centrosymmetric itinerant magnets by taking the Kondo-lattice model on a triangular lattice, which is known to exhibit two types of skyrmion lattices with different magnetic topological charges of $|N_{\rm sk}|$=1 and $|N_{\rm sk}|$=2. Our numerical simulations reveal that intense excitation of a resonance mode with circularly polarized microwave field can switch the magnetic topology, i.e., from the skyrmion lattice with $|N_{\rm sk}|$=1 to another skyrmion lattice with $|N_{\rm sk}|$=2 or to a nontopological magnetic order with $|N_{\rm sk}|$=0 depending on the microwave frequency. This magnetic-topology switching shows various distinct behaviors, that is, deterministic irreversible switching, probabilistic irreversible switching, and temporally random fluctuation depending on the microwave frequency and the strength of external magnetic field, variety of which is attributable to different energy landscapes in the dynamical regime. The obtained results are also discussed in the light of time-evolution equations based on an effective model derived using perturbation expansions.
\end{abstract}
\maketitle

\section{Introduction}
\begin{figure}[tbh]
\centering
\includegraphics[scale=0.4]{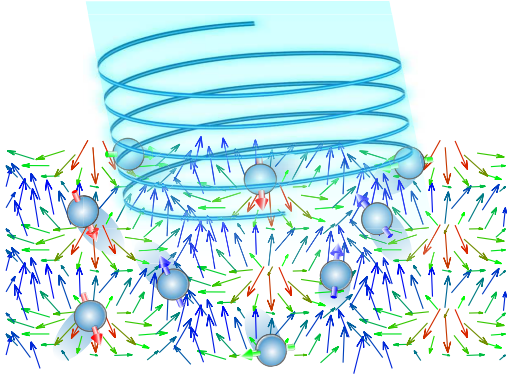}
\caption{Schematics of a skyrmion lattice in the itinerant magnet irradiated with circularly polarized microwave field. The thin solid arrows represent local magnetizations, whereas the light blue spheres with thick solid arrows represent itinerant electrons with spins.}
\label{Fig01}
\end{figure}
Topological magnetisms exemplified by several types of skyrmions~\cite{Muhlbauer2009,Yu2010,Heinze2011,Kezsmarki2015,Yu2014,Nayak2017}, merons~\cite{Yu2018}, hedgehogs~\cite{Milde2013,Schutte2014,Kanazawa2012,Fujishiro2019,Ishiwata2020}, and hopfions~\cite{LiuY2018} are currently attracting enormous research interest from the viewpoints of both fundamental sciences and potential applications~\cite{Seki2016,Nagaosa2013,Gobel2021,Fert2013,Finocchio2016,Fert2017,Everschor2019,Tokura2021}. These nontrivial magnetic textures with spatially modulated magnetization are often caused by the Dzyaloshinskii-Moriya interactions~\cite{Dzyaloshinsky1958,Moriya1960a,Moriya1960b}, which have a relativistic origin and are active when the system has no spatial inversion symmetry. Therefore, the above topological magnetic textures are usually hosted in magnets having non-centrosymmetric crystal structures or magnetic heterostructures having interfaces. In such magnets, structural symmetries and anisotropies determine the way of magnetization alignment in magnetic textures through governing the spatial configuration of Dzyaloshinskii-Moriya vectors~\cite{Bogdanov1989,Bogdanov1994,Tanaka2020}. Consequently, the helicity, chirality, and vorticity of the magnetic textures are inherently fixed and thus are not variable~\cite{Nagaosa2013}.

Exchange coupling between itinerant electrons and local magnetizations is another source of topological magnetisms~\cite{Hayami2021R,Hayami2014a,Ozawa2016,Ozawa2017a,Hayami2017,Gobel2017,Wang2020,Hayami2020a,Hayami2021a,Hayami2021b,Nakazawa2019,Hayami2020b,Hayami2018,Okada2018,Okumura2020,Shimizu2021a}. Recent theoretical studies have predicted possible emergence of topological magnetisms such as skyrmion lattices~\cite{Ozawa2016,Ozawa2017a,Hayami2017,Gobel2017,Wang2020,Hayami2020a,Hayami2021a,Hayami2021b}, meron lattices~\cite{Hayami2021R,Hayami2021b}, and hedgehog lattices~\cite{Hayami2021R,Okumura2020,Shimizu2021a} in the Kondo-lattice model and its effective model which describe localized magnetizations on a lattice coupled to itinerant electron spins via (anti)ferromagnetic exchange interactions. In such magnets, spatial modulation of magnetization is caused by effective long-range interactions among local magnetizations mediated by conduction electrons [e.g., the Ruderman-Kittel-Kasuya-Yosida (RKKY) interactions~\cite{Ruderman1954,Kasuya1956,Yosida1957}]. Therefore, in the case of the Kondo-lattice magnets, even the centrosymmetric systems with spatial inversion symmetry can host topological magnetisms, despite the Dzyaloshinskii-Moriya interactions are absent~\cite{Hayami2021R,Ozawa2016,Ozawa2017a,Hayami2017,Gobel2017,Wang2020,Hayami2020a,Hayami2021a,Hayami2021b,Nakazawa2019,Hayami2020b}. These magnetic textures are superpositions of magnetic helices with multiple propagation vectors ($\bm Q$ vectors), which are determined by the effective electron-mediated interactions among magnetizations governed by multiple nesting vectors of the Fermi surfaces. This fact enables us to have a variety of topological magnetic textures and their controllability via material variations or by tuning material parameters~\cite{Shimizu2021b}. Recent experiments have indeed observed three-dimensional hedgehog lattices in SrFeO$_3$~\cite{Ishiwata2020}, triangular skyrmion lattices in Gd$_2$PdSi$_3$~\cite{Kurumaji2019,Hirschberger2020a,Hirschberger2020b,Nomoto2020} and Gd$_3$Ru$_4$Al$_{12}$~\cite{Hirschberger2019,Hirschberger2021}, and square skyrmion lattices in GdRu$_2$Si$_2$~\cite{Khanh2020,Yasui2020}, all of which are itinerant magnets with centrosymmetric crystal structure.

In addition to their rich variety, topological magnetic textures in centrosymmetric metallic magnets have an interesting feature, that is, several degrees of freedom such as helicity and vorticity remain to be unfrozen because the Dzyaloshinskii-Moriya interaction is absent. Thereby, their low-energy excitations and continuous variations are possible, which provide us a unique opportunity to control and switch the magnetic topology with an external stimulus such as static magnetic field, electric currents~\cite{SZLin2013}, and microwave magnetic field. It has been revealed that topological magnetisms such as skyrmions and skyrmion tubes in chiral magnets have peculiar collective modes or spin-wave modes at microwave frequencies~\cite{Mochizuki2012,Schwarze2015,LinSZ2014,McKeever2019,LinSZ2019,Seki2020}, and they have turned out to cause interesting physical phenomena and potential device functions~\cite{Mochizuki2015,Garst2017,Lonsky2020,Mochizuki2013,Okamura2013,MaF2015,WangK2020,WangW2015,Ikka2018,Takeuchi2018,SongC2019,YuanHY2019,Koide2019,Takeuchi2019,Miyake2020,XingX2020,Ser2020}. We expect that the topological magnetisms in centrosymmetric magnets can also host interesting dynamical phenomena and functionalities associated with their microwave-active collective modes.

In this paper, we theoretically demonstrate that magnetic topology in a centrosymmetric itinerant magnet can be switched dynamically by application of circularly polarized microwave field by taking the Kondo-lattice model on a triangular lattice as an example. This model is known to exhibit two distinct skyrmion-lattice phases [Fig.\ref{Fig01}] with different skyrmion numbers or magnetic topological charges $|N_{\rm sk}|$=1 and $|N_{\rm sk}|$=2~\cite{Ozawa2017a}. We numerically simulate microwave-driven magnetization dynamics in this model using a combined technique of the micromagnetic simulation and the kernel polynomial method. We first find that a single spin-wave mode can be activated by an in-plane microwave field in all the three phases. We then demonstrate that by intensely activating this spin-wave mode with circularly polarized microwave field, the skyrmion lattice with $|N_{\rm sk}|$=1 can be switched to that with $|N_{\rm sk}|$=2 or a nontopological magnetic state of $N_{\rm sk}$=0 depending on the microwave frequency. During the switching processes, we observe emergent topological magnetic patterns characterized by half-integer skyrmion numbers of $|N_{\rm sk}|$=1/2 and $|N_{\rm sk}|$=3/2 (i.e., meron lattices) as transient states. We examine such dynamical transitions for various initial magnetic configurations in equilibrium and find several different behaviors, that is, deterministic irreversible switching, probabilistic irreversible switching, and temporally random fluctuations under continuous microwave irradiation. This variety of behaviors is attributable to difference of the energy landscape in the dynamical regime. We also discuss the obtained results on the basis of an effective model derived using perturbation expansions~\cite{Hayami2017}. Note that magnetic frustration is another important mechanism to realize topological magnetism in centrosymmetric magnets. The phenomena revealed in this work can also be expected in the frustrated systems although fine tuning of the exchange interactions is required to produce topological magnetic textures of frustration origin~\cite{Okubo2012,Kamiya2014,Leonov2015,LinSZ2016,Hayami2016a,Hayami2016b,LinSZ2018,Lohani2019,WangZ2021}.

\section{Model}
\subsection{Kondo-Lattice Model}
\begin{figure*}[tbh]
\centering
\includegraphics[scale=1.0]{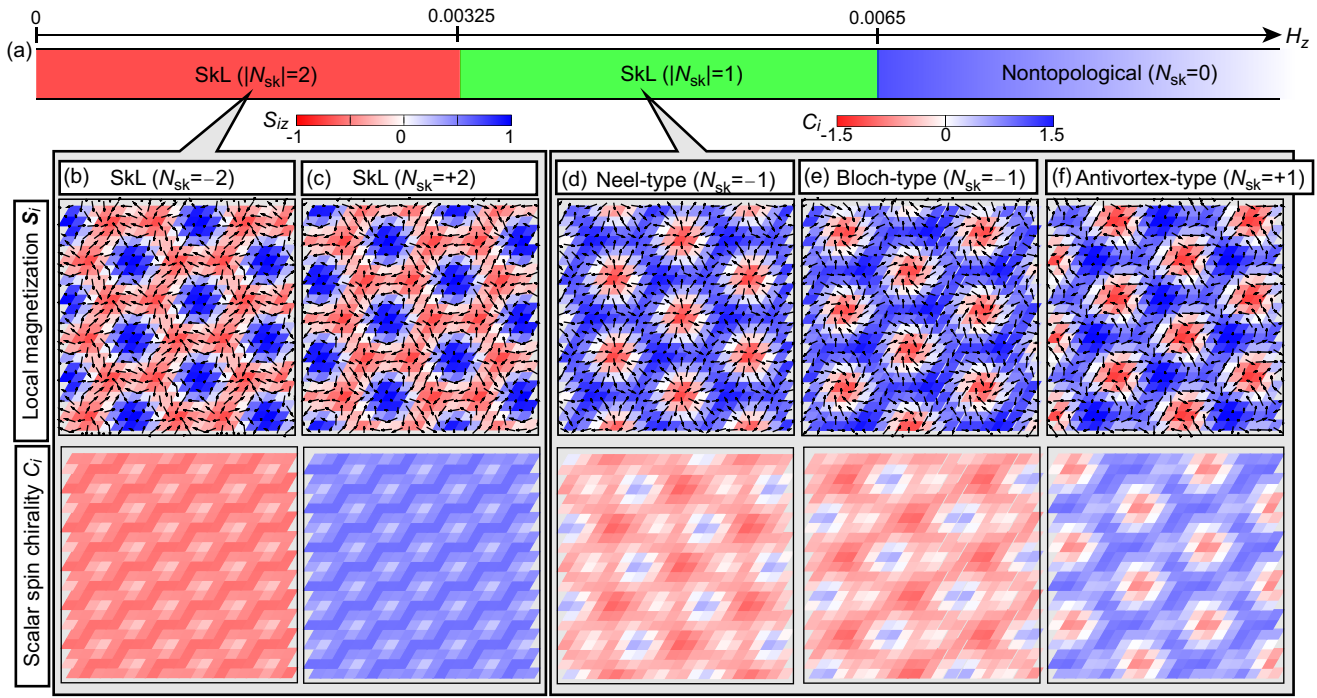}
\caption{(a) Ground-state phase diagram of the Kondo-lattice model in Eq.~(\ref{Hamlt}) on the triangular lattice as a function of static magnetic field $H_z$ perpendicular to the lattice plane when the microwave magnetic field is absent (i.e., $\bm H(t)=0$)~\cite{Ozawa2017a}. The model parameters are set to be $t_1=-1$, $t_3=0.85$, $J_{\rm K}=-0.5$, and $\mu=-3.5$. Successive two phase transitions among three phases, i.e., the skyrmion-lattice (SkL) phase with $|N_{\rm sk}|=2$, another SkL phase with $|N_{\rm sk}|=1$, and the nontopological phase with $N_{\rm sk}=0$ take place at $H_z=0.00325$ and $H_z=0.0065$. Several degrees of freedom of magnetic textures such as helicity and vorticity are not frozen in the centrosymmetric system without Dzyaloshinskii-Moriya interactions, which results in infinite degeneracy of magnetic textures (see text). (b), (c) Spatial profiles of local magnetizations (upper panels) and local scalar spin chiralities (lower panels) of two degenerate SkLs with $|N_{\rm sk}|=2$, i.e., (b) SkL with $N_{\rm sk}=-2$ and (c) SkL with $N_{\rm sk}=+2$. (d)-(f) Those of SkLs with $|N_{\rm sk}|=1$, i.e., (d) N\'{e}el-type SkL with $N_{\rm sk}=-1$, (e) Bloch-type SkL with $N_{\rm sk}=-1$, and (f) antivortex-type SkL with $N_{\rm sk}=+1$.}
\label{Fig02}
\end{figure*}
We consider the Kondo-lattice model on a triangular lattice. The Hamiltonian is given by
\begin{align}
\mathcal{H}=\mathcal{H}_{\rm KL} + \mathcal{H}_{\rm Zeeman},
\label{Hamlt}
\end{align}
with
\begin{align}
&\mathcal{H}_{\rm KL}=
\sum_{ij\sigma}t_{i,j} \hat{c}^\dagger_{i\sigma}\hat{c}_{j\sigma} 
+J_{\rm K}\sum_{i,\sigma,\sigma'}\hat{c}^\dagger_{i\sigma}{\bm \sigma}_{\sigma\sigma'}
\hat{c}_{i\sigma'}\cdot{\bm S}_i,
\label{eq:KLM} \\
&\mathcal{H}_{\rm Zeeman}=
-\sum_i \left[\bm H_{\rm ext} + \bm H(t) \right]\cdot{\bm S}_i.
\label{Zeeman}
\end{align}
Here $\hat{c}^\dagger_{i\sigma}$ $(\hat{c}_{i\sigma})$ denotes the creation (annihilation) operator of an itinerant electron with spin $\sigma(=\uparrow,\downarrow)$ on the $i$th site, and ${\bm S}_i$ denotes localized magnetization on the $i$th site. The first term of $\mathcal{H}_{\rm KL}$ represents kinetic energies of itinerant electrons where the nearest-neighbor hopping $t_1$ and the third-nearest-neighbor hopping $t_3$ are set to be $t_1=-1$ and $t_3=0.85$, respectively. The second term of $\mathcal{H}_{\rm KL}$ represents the exchange coupling between itinerant electron spins and local magnetizations where the coupling constant is set to be $J_{\rm K}=-0.5$. The term $\mathcal{H}_{\rm Zeeman}$ represents the Zeeman couplings associated with both a static external magnetic field 
$\bm H_{\rm ext}=(0,0,H_z)$ and a time-dependent magnetic field $\bm H(t)$ acting on the local magnetizations. For the time-dependent field $\bm H(t)$, we consider a short-period pulse or circularly polarized microwave field in the present study. Note that we neglect the coupling between the magnetic fields and the itinerant electrons. In fact, we have examined the effects of the coupling and have found that consideration of the coupling does not alter the results even quantitatively. We set the chemical potential $\mu=-3.5$, which corresponds to the electron filling of 0.2 approximately. The above parameter values are the same as those used in the previous work~\cite{Ozawa2017a}. 

This model is known to exhibit various magnetic orders including topological ones as superpositions of three magnetic helices. The propagation vectors of the three helices are $\bm Q_1=(\pi/3, 0)$, $\bm Q_2=\hat{R}(2\pi/3)\bm Q_1$, and $\bm Q_3=\hat{R}(4\pi/3)\bm Q_1$ where $\hat{R}(\phi)$ is an operator to rotate the vector by the angle $\phi$ around the $z$-axis. A ground-state phase diagram was studied in Ref.~\cite{Ozawa2017a} as a function of the strength of external magnetic field $H_z$ when the microwave field $\bm H(t)$ is absent [Fig.~\ref{Fig02}(a)]. We find that three magnetic phases, i.e., a skyrmion-lattice phase with $|N_{\rm sk}|=2$, another skyrmion-lattice phase with $|N_{\rm sk}|=1$ and a nontopological phase with $N_{\rm sk}=0$ successively emerge as $H_z$ increases. Note that the skyrmion-lattice with $|N_{\rm sk}|=2$ emerges even at $H_z=0$ in striking contrast to the case of the Dzyaloshinskii-Moriya magnets in which topological magnetisms usually appear in the presence of an external magnetic field. More importantly, several degrees of freedom remain unfrozen in the present system with spatial inversion symmetry. For example, the helicity and the signs of vorticity are not frozen for the skyrmion-lattice phase with $|N_{\rm sk}|=1$ in the presence of the spatial inversion symmetry. Thereby, the magnetic structures in the present system have infinite degeneracy [see Figs.~\ref{Fig02}(b)-(f)].

It should be mentioned that the present triangular-lattice system is favorable for the emergence of the skyrmion lattices with a higher topological number of $|N_{\rm sk}|=2$. In the centrosymmetric Kondo-lattice system, the skyrmion lattices emerge as a superposition of three spiral or sinusoidal states of local magnetizations, which are stabilized by the long-ranged and frustrated RKKY interactions. The RKKY interactions originate from the coupling between conduction electrons and local magnetizations and thus are governed by the electronic structure of conduction electrons, e.g., the Fermi-surface geometry and the density of states. Consequently, the modulation vectors $\bm Q_\nu$ determined by the RKKY interactions also depend on the Fermi-surface geometry. More specifically, these modulation vectors correspond to nesting vectors of the Fermi surface(s) and, thereby, reflect the symmetry of the lattice structure. In addition, it has turned out that the skyrmion lattices with $|N_{\rm sk}|=2$ requires, at least, three magnetization spiral or sinusoidal states, while the skyrmion lattices with $|N_{\rm sk}|=1$ can be produced only with two magnetization spiral or sinusoidal states. Hence, the triangular lattices and the Kagome lattices with triangular or hexagonal symmetries have more opportunity to host the skyrmion lattices with $|N_{\rm sk}|=2$ as compared to the simple square lattices.

\subsection{Global Symmetry in Skyrmion Lattices}
Let us discuss the degeneracy of topological magnetic textures in the present centrosymmetric system in more detail. We first consider the cases without external magnetic field. The local magnetizations $\bm S_i$ for the skyrmion-lattice phases with $|N_{\rm sk}|=1$ and $|N_{\rm sk}|=2$ are given by,
\begin{align}
{\bm S}_i^{N_{\rm Sk}=\pm 1} &\propto \sum_{\nu=1}^3
\left(
\begin{array}{c}
\sin\mathcal{Q}_\nu\cos\phi_\nu \\
\lambda_{\rm v}\sin\mathcal{Q}_\nu\sin\phi_\nu \\
\cos\mathcal{Q}_\nu
\end{array}
\right),
\\
{\bm S}_i^{N_{\rm Sk}=-2}&\propto
\left(
\begin{array}{c}
\cos\mathcal{Q}_1 \\
\cos\mathcal{Q}_2 \\
\cos\mathcal{Q}_3
\end{array}
\right),
\\
{\bm S}_i^{N_{\rm Sk}=+2}&\propto
\left(
\begin{array}{c}
\cos\mathcal{Q}_1 \\
\cos\mathcal{Q}_3 \\
\cos\mathcal{Q}_2
\end{array}
\right),
\end{align}
where 
\begin{align}
\mathcal{Q}_\nu={\bm Q}_\nu\cdot{\bm r}_i+\theta_\nu,
\quad
\phi_\nu=2\pi(\nu-1)/3.
\end{align}
Here the angles $\theta_\nu$ ($\nu$=1,2,3) represent phase shifts, and $\Theta=\sum_{\nu=1}^3\theta_\nu$ is an internal degree of freedom called phason~\cite{Nakazawa2019,Hayami2020b}. The variable $\lambda_{\rm v}(=\pm 1)$ is called vorticity, and $\lambda_{\rm v}=+1 (-1)$ corresponds to $N_{\rm sk}=-1 (+1)$. Because the variation of $\Theta$ is accompanied by a change in energy, we set $\theta_\nu$=0 hereafter. These formulae represent that the skyrmion-lattice with $|N_{\rm sk}|=1$ is a superposition of the three helices, while that with $|N_{\rm sk}|=2$ is a superposition of the three cosine waves. 

These two skyrmion lattices break different symmetries. Specifically, the skyrmion lattices with $|N_{\rm sk}|=1$ break the U(1) symmetry associated with the in-plane rotational invariance and thus have a degree of freedom called helicity. The variation of magnetic texture upon the helicity shift by $\phi_1$ is given, e.g., by, 
\begin{align}
{\bm S}_i^{N_{\rm Sk}=\pm 1} 
&\propto \hat{R}_z(\phi_1) \sum_{\nu=1}^3 
\left(
\begin{array}{c}
\sin\mathcal{Q}_\nu\cos\phi_\nu \\
\lambda_{\rm v}\sin\mathcal{Q}_\nu\sin\phi_\nu \\
\cos\mathcal{Q}_\nu
\end{array}
\right) 
\label{eq:U1rot}
\end{align}
where $\hat{R}_\gamma(\varphi)$ is an operator to rotate the vector by the angle $\varphi$ around the $\gamma$-axis. On the other hand, the skyrmion lattices with $|N_{\rm sk}|=2$ break the SO(3) symmetry. The variation of magnetic texture upon the SO(3)-invariant rotational operations by angles $\theta_2$ and $\phi_2$ is given, e.g., by, 
\begin{align}
{\bm S}_i^{N_{\rm Sk}=-2} &\propto \hat{R}_x(\theta_2)\hat{R}_z(\phi_2)
\left(
\begin{array}{c}
\cos\mathcal{Q}_1 \\
\cos\mathcal{Q}_2 \\
\cos\mathcal{Q}_3
\end{array}
\right).
\label{eq:SO3rot}
\end{align}
Note that the energy of the skyrmion lattice with $|N_{\rm sk}|=1$ does not change upon the variation of $\phi_1$, whereas that with $|N_{\rm sk}|=2$ does not change upon the variations of $\theta_2$ and $\phi_2$.

Then we consider the effects of external magnetic field. When a magnetic field is applied, some of the symmetries mentioned in the above discussion would be violated. The U(1) symmetry around the $z$-axis in the skyrmion lattice with $|N_{\rm sk}|=1$ becomes absent when the external magnetic field has in-plane components. Therefore, the magnetic structures of the $|N_{\rm sk}|=1$ skyrmion lattice do not have any global symmetries when irradiated with circularly polarized microwave field. In addition, the SO(3) symmetry in the skyrmion lattice with $|N_{\rm sk}|=2$ is partially violated by the external magnetic field, whereas the U(1) symmetry around the magnetic field remains. More specifically, the $|N_{\rm sk}|=2$ skyrmion lattice has the U(1) symmetry around the total magnetic field at every moment, and thus its symmetry axis temporally varies. We also note that a magnetic field also modulates a degree of freedom associated with phasons $\Theta$.

\section{Method}
We simulate time evolution of the local magnetizations in the Kondo-lattice system by numerically solving the Landau-Lifshitz-Gilbert (LLG) equation. The LLG equation is given by,
\begin{eqnarray}
\frac{d{\bm S}_i}{dt} = -{\bm S}_i\times{\bm H}^{\rm eff}_i + \frac{\alpha_{\rm G}}{S}{\bm S}_i\times\frac{d{\bm S}_i}{dt},
\label{LLG}
\end{eqnarray}
where $\alpha_{\rm G}(=0.05)$ is the dimensionless Gilbert-damping constant, and $S(=1)$ is the saturation magnetization. The effective magnetic field ${\bm H}^{\rm eff}_i$ acting on the local magnetization at the $i$th site is calculated by,
\begin{eqnarray}
{\bm H}^{\rm eff}_i = -\frac{\partial\Omega}{\partial{\bm S}_i} + {\bm H}_{\rm ext} + {\bm H}(t).
\end{eqnarray}
Here $\Omega$ is the grand canonical potential of $\mathcal{H}_{\rm KL}$, which is given by,
\begin{eqnarray}
\Omega = \int \rho(\varepsilon,\{{\bm S}_i\})F(\varepsilon-\mu)\;d\varepsilon, 
\end{eqnarray}
with
\begin{eqnarray}
\rho(\varepsilon,\{{\bm S}_i\}) = \frac{1}{2N}\sum_{k=1}^{2N}\delta(\varepsilon-\varepsilon_k(\{{\bm S}_i\})).
\end{eqnarray}
Here $F(\varepsilon-\mu)$ is the free energy of the system, and $\rho(\varepsilon,\{{\bm S}_i\})$ is the density of state of conduction electrons for a given set of the local magnetizations $\{{\bm S}_i\}$. To calculate $\Omega$ and its magnetization-derivatives $\partial\Omega/\partial{\bm S}_i$, we adopt the kernel polynomial method, which is based on the Chebyshev polynomial expansion of $\Omega$ and the automatic differentiation~\cite{Motome1999,Weisse2006,Barros2013,Wang2016,Ozawa2017b,Chern2018,Wang2018}. 

All the simulations are performed at zero temperature with no thermal fluctuations in order to demonstrate that the microwave application can solely induce the magnetic topological switching and to capture the physics of this field-induced phenomenon. For the simulations, a lattice with $N=36^2$ sites, on which periodic boundary conditions are imposed, is adopted. We use 324 correlated random vectors \cite{Wang2018,Tang2012} for simulating relaxation dynamics to obtain initial magnetic configurations through minimizing the energy, whereas we adopt complete orthonormal basis states for the simulations of microwave-induced dynamics. We use Chebyshev polynomials up to the 2000th-order for the expansion of $\Omega$ and adopt the fourth-order Runge-Kutta method with a time slice of $\Delta t=4$ to solve the LLG equation in Eq.~(\ref{LLG}). The spatiotemporal dynamics of local magnetizations $\bm S_i$ and local scalar spin chiralities $C_i$ are computed. The spin chirality $C_i$ is calculated by,
\begin{align}
C_i=
 {\bm S}_i \cdot{\bm S}_{i+\hat{a}} \times {\bm S}_{i+\hat{a}+\hat{b}}
+{\bm S}_i \cdot{\bm S}_{i+\hat{a}+\hat{b}} \times {\bm S}_{i+\hat{b}},
\end{align}
where $\hat{a}$ and $\hat{b}$ are the primitive lattice vectors of triangular lattice. We also compute time profiles of the net magnetization $\bm S$ and the skyrmion number $N_{\rm sk}$, which are respectively calculated by~\cite{Kim2020},
\begin{align}
&\bm S=\frac{1}{N} \sum_{i=1}^N \bm S_i, \\
&N_{\rm sk} = \frac{1}{4\pi N_{\rm m}}\sum_{i=1}^N
\nonumber \\
&\left[ 2\tan^{-1}\left(
\frac{{\bm S}_i \cdot {\bm S}_{i+\hat{a}} \times {\bm S}_{i+\hat{a}+\hat{b}}}
{1+ {\bm S}_i \cdot {\bm S}_{i+\hat{a}} + {\bm S}_{i+\hat{a}} \cdot {\bm S}_{i+\hat{a}+\hat{b}}
+{\bm S}_{i+\hat{a}+\hat{b}} \cdot {\bm S}_i} 
\right)\right.
\nonumber \\
&+ \left.
2\tan^{-1}\left(
\frac{{\bm S}_i \cdot {\bm S}_{i+\hat{a}+\hat{b}} \times {\bm S}_{i+\hat{b}}}
{1+ {\bm S}_i \cdot {\bm S}_{i+\hat{a}+\hat{b}} + {\bm S}_{i+\hat{a}+\hat{b}} \cdot {\bm S}_{i+\hat{b}}+{\bm S}_{i+\hat{b}} \cdot {\bm S}_i}
\right)
\right].
\end{align}
Here $N_{\rm m}(=27)$ is the number of magnetic unit cells where one unit cell contains 48 sites. Note that the sizes of magnetic unit cells are common for all the magnetic patterns which appear in the present simulations because they are all constituted with three magnetic helices with the same wave vectors $\bm Q_\nu$ ($\nu=1,2,3$) determined by the Fermi-surface nesting.

\section{Results}
\subsection{Spin-Wave Modes}
\begin{figure}[tbh]
\centering
\includegraphics[scale=1.0]{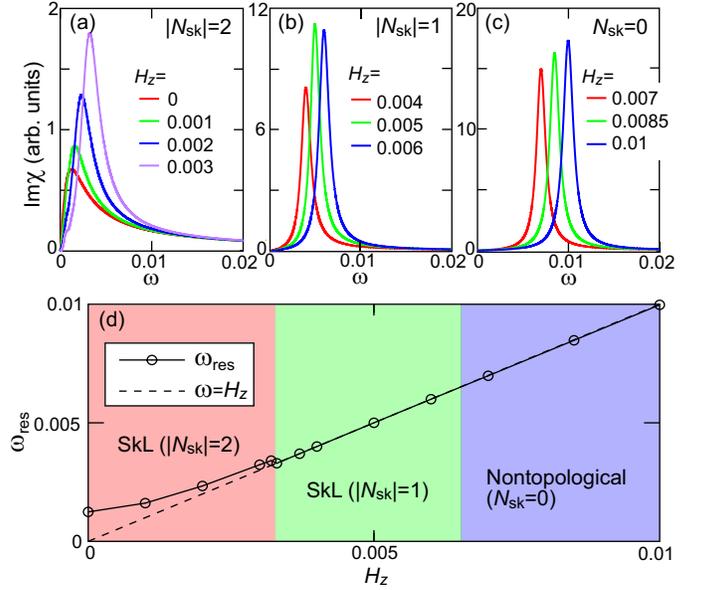}
\caption{(a)-(c) Microwave-absorption spectra for in-plane microwave magnetic fields in respective magnetic phases, i.e., (a) the skyrmion-lattice phase with $|N_{\rm sk}|=2$, (b) the skyrmion-lattice phase with $|N_{\rm sk}|=1$, and (c) the nontopological phase with $N_{\rm sk}=0$. (d) $H_z$-dependence of resonance frequency $\omega_{\rm res}$ for spin-wave modes active to an in-plane microwave field (corresponding to peak positions of the microwave-absorption spectra).}
\label{Fig03}
\end{figure}
First, we study microwave-active resonance modes in each magnetic phase by numerically calculating the dynamical magnetic susceptibilities,
\begin{eqnarray}
\chi_\gamma(\omega) =
\frac{\Delta S_\gamma(\omega)}{H_\gamma(\omega)} \quad\quad
(\gamma=x, y, z),
\end{eqnarray}
where $H_\gamma(\omega)$ and $\Delta M_\gamma(\omega)$ are Fourier components of the time-dependent magnetic field $\bm H(t)$ and those of the time-profile of total magnetization $\Delta \bm S(t)=\bm S(t)-\bm S(0)$. Here we particularly focus on the resonance modes active to the in-plane polarized microwave field and thus set $\gamma=x$. To calculate these quantities, we adopt a spatially uniform short-time pulse of magnetic field for $\bm H(t)$, which is given by,
\begin{eqnarray}
\bm H(t)=\left\{
\begin{aligned}
& (H_{\rm pulse},0,0) \quad & 0 \le t \le 1 \\
& 0 \quad & {\rm others}
\end{aligned}
\right.
\end{eqnarray}
where $t=(t_1/\hbar)\tau$ is the dimensionless time with $\tau$ and $t_1$ respectively being the real time and the nearest-neighbor hopping integral. We compute time evolutions of local magnetizations $\bm S_i(t)$ and their sum $\bm S(t)$ after applying this pulse field to the system. The usage of the short-time pulse is advantageous because the Fourier components $H_\gamma(\omega)$ become constant being independent of $\omega$ up to the first order of $\omega \Delta t$ for a sufficiently short duration $\Delta t$ (i.e., $\omega \Delta t \ll 1$). The Fourier components are calculated as 
\begin{eqnarray}
H_\gamma(\omega)&=&\int_0^{\Delta t}\;H_{\rm pulse} e^{i\omega t}dt
=\frac{H_{\rm pulse}}{i\omega}\left(e^{i\omega \Delta t}-1 \right)
\nonumber \\
&\sim&H_{\rm pulse} \Delta t.
\end{eqnarray}
As a result, we obtain the relationship $\chi_\gamma(\omega) \propto \Delta M_\gamma(\omega)$.

In Figs.~\ref{Fig03}(a)-(c), we present the calculated microwave absorption spectra, i.e., imaginary part of the dynamical magnetic susceptibility Im$\chi_x$, for (a) the skyrmion-lattice phase with $|N_{\rm sk}|$=2, (b) the skyrmion-lattice phase with $|N_{\rm sk}|$=1, and (c) the nontopological phase with $N_{\rm sk}$=0. Each of the spectra has a single peak, indicating the existence of a single resonance mode in each phase. The mode in the skyrmion-lattice phase with $|N_{\rm sk}|$=1 is a rotational mode in which all the skyrmions constituting the skyrmion lattice rotate uniformly in the counterclockwise fashion. It is known that skyrmion lattices in the Dzyaloshinskii-Moriya magnets without inversion symmetry exhibit two rotation modes with opposite rotation senses (i.e., counterclockwise and clockwise) at different frequencies~\cite{Mochizuki2012}, whereas skyrmion lattices stabilized by frustrated exchange interactions in centrosymmetric Heisenberg magnets exhibit a counterclockwise mode only~\cite{Leonov2015}. The situation in our centrosymmetric metallic magnets with the RKKY interactions resembles the latter case. Figure~\ref{Fig03}(d) presents the resonance frequency $\omega_{\rm res}$ as a function of external magnetic field $H_z$. We find that a relation $\omega=H_z$ (i.e., $\omega=g\mu_{\rm B}H_z/\hbar$ in dimensionfull units) holds in the phases with $|N_{\rm sk}|$=1 and $N_{\rm sk}$=0.

\subsection{Microwave-Induced Dynamics}
\begin{figure*}[tbh]
\begin{center}
\includegraphics[scale=1.0]{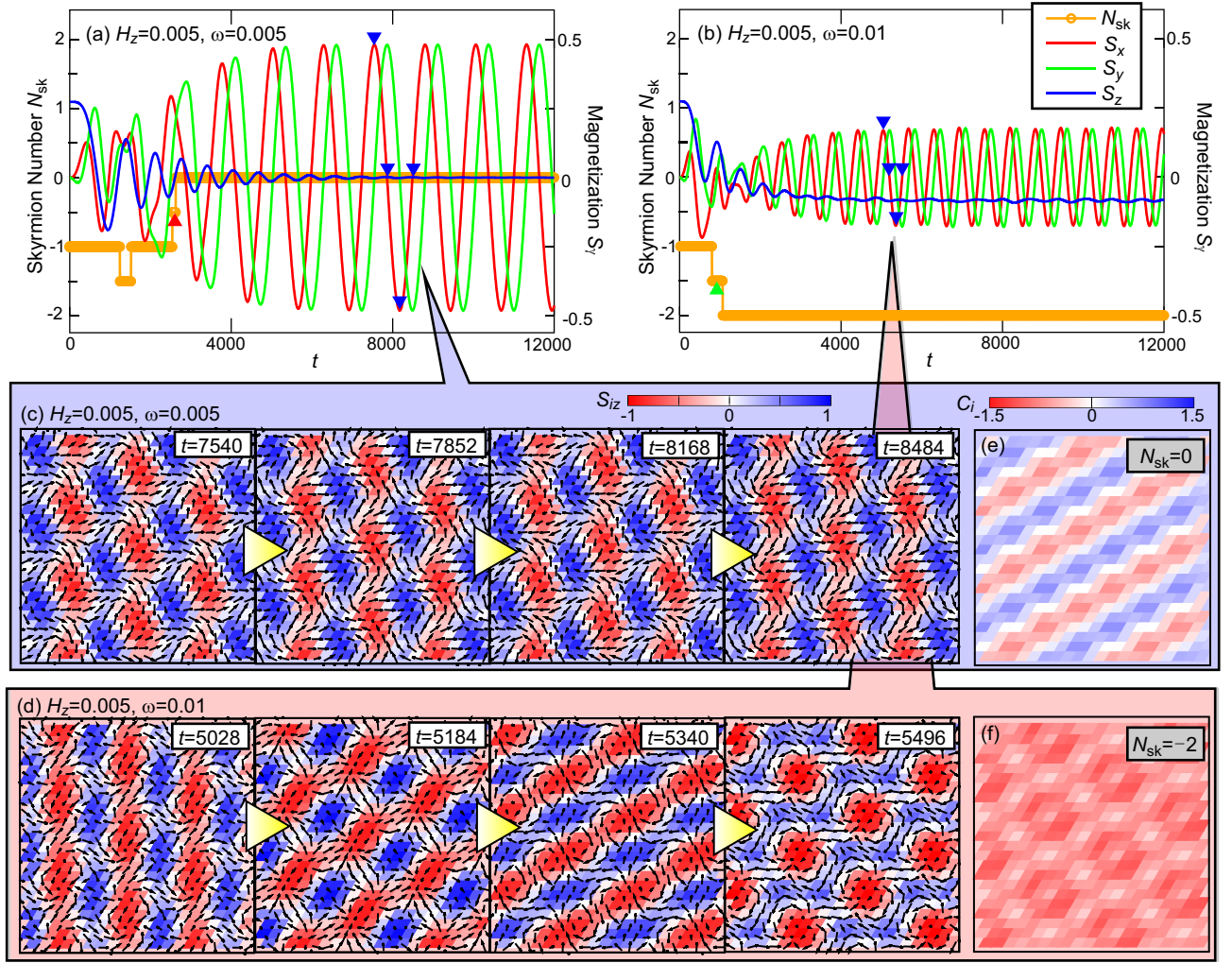}
\caption{Simulated time profiles of the net magnetization $\bm S=(S_x,S_y,S_z)$ and the skrmion number $N_{\rm sk}$ in the Kondo-lattice model under irradiation with circularly polarized microwave field with (a) $\omega$=0.005 and (b) $\omega$=0.01 when $H_z=0.005$. For both cases, we start with a skyrmion-lattice configuration with $N_{\rm sk}=-1$ as an initial state, which is the ground state at $H_z=0.005$. (c),(d) Snapshots of the temporally varying spatial profiles of local magnetizations $\bm S_i$ in the microwave-induced nonequilibrium steady phases, i.e., (c) the nontopological phase $N_{\rm sk}=0$ and (d) the skyrmion-lattice phase with $N_{\rm sk}=-2$ at selected moments indicated by inverted triangles in (a) and (b), respectively. (e), (f) Spatial profiles of the local scalar spin chiralities $C_i$ in each nonequilibrium steady phase, which do not change temporally.}
\label{Fig04}
\end{center}
\end{figure*}
\begin{figure}[tbh]
\centering
\includegraphics[scale=1.0]{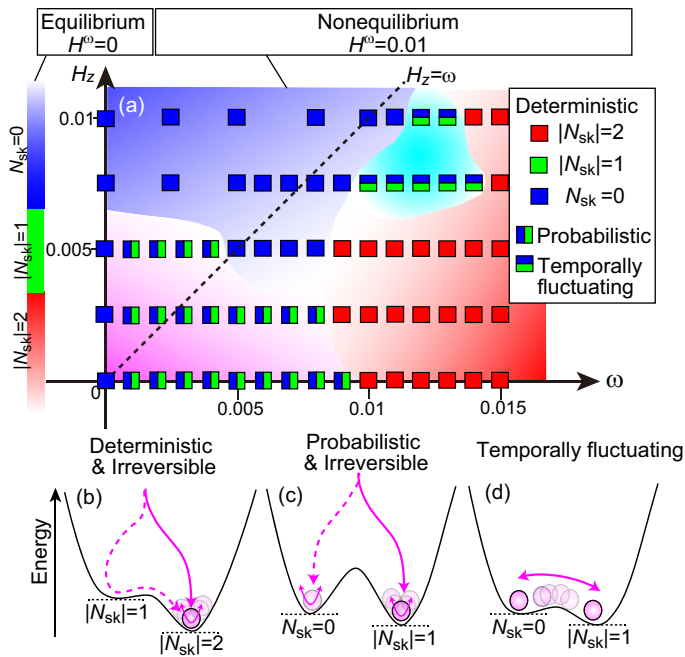}
\caption{Nonequilibrium phase diagram in plane of the microwave frequency $\omega$ and the strength of static perpendicular magnetic field $H_z$ under irradiation with circularly polarized microwave field of $H^\omega$=0.01. The Gilbert-damping coefficient is set to be $\alpha_{\rm G}$=0.05 for the simulations. Equilibrium phase diagram in the absence of microwave field (i.e., $H^\omega$=0) is also presented at the left end. (b)-(d) Schematics of the energy landscapes for different switching behaviors, i.e., (b) the deterministic irreversible switching, (c) the probabilistic irreversible switching, and (d) the temporally random fluctuation.}
\label{Fig05}
\end{figure}
Next we simulate the magnetization dynamics under irradiation with circularly polarized microwave field, which is given by,
\begin{equation}
{\bm H}(t) = H^\omega\beta(t)(\cos\omega t,\sin\omega t,0),
\label{eqn:CPMW}
\end{equation}
with
\begin{equation}
\beta(t)=\tanh\left(\frac{t}{\tau_{\rm d}}\right).
\label{eqn:CPMW2}
\end{equation}
Here the time-dependent prefactor $\beta(t)$ with $\tau_{\rm d}=2\pi/\omega$ is introduced to avoid unexpected artifacts due to impact-force effects through gradually increasing the microwave amplitude.


Figures~\ref{Fig04}(a) and (b) present simulated time profiles of the net magnetization $\bm S=(S_x,S_y,S_z)$ and the skrmion number $N_{\rm sk}$ in a system irradiated with circularly polarized microwave field with (a) $\omega$=0.005 and (b) $\omega$=0.01 when $H_z=0.005$. We start the simulation with a skyrmion-lattice configuration with $N_{\rm sk}=-1$ as an initial state for both cases, which is the ground state at $H_z=0.005$. We observe a microwave-induced switching of the magnetic topology from $N_{\rm sk}=-1$ to $N_{\rm sk}=0$ in Fig.~\ref{Fig04}(a), whereas that from $N_{\rm sk}=-1$ to $N_{\rm sk}=-2$ in Fig.~\ref{Fig04}(b). The phase with $N_{\rm sk}=-1$ and the phase with $N_{\rm sk}=0$ in respective cases appear as nonequilibrium steady phases where the net magnetizations show steady oscillations. 

Although the skyrmion number $N_{\rm sk}$ is constant, the spatial magnetic configurations in these nonequilibrium steady phases under irradiation with microwave field vary periodically in time. Figures~\ref{Fig04}(c) and (d) present a series of snapshots of the temporally varying local magnetizations $\bm S_i$ for (c) the nontopological phase $N_{\rm sk}=0$ and (d) the skyrmion-lattice phase with $N_{\rm sk}=-2$ under irradiation with the microwave field at selected moments. In fact, as long as the skyrmion number is constant, these magnetic configurations are connected to each other by certain rotational operations. For example, the four magnetic configurations with $N_{\rm sk}=-2$ shown in Fig.~\ref{Fig04}(d) are all connected to each other via the SO(3)-invariant rotational operations by angles $\theta_2$ and $\phi_2$ represented by Eq.~(\ref{eq:SO3rot}). Note that the local scalar spin chiralities $C_i$ are time-independent in contrast to the local magnetizations $\bm S_i$. In Figs.~\ref{Fig04}(e) and (f), we present the spatial profiles of $C_i$ in the microwave-induced $N_{\rm sk}=0$ and $N_{\rm sk}=-2$ phases, respectively, which do not change temporally.

Then we study nonequilibrium magnetic phases after sufficient a duration of the microwave irradiation by varying the microwave frequency $\omega$ and the applied static magnetic field $H_z$. Figure~\ref{Fig05} presents the obtained nonequilibrium phase diagram in plane of $\omega$ and $H_z$ for a system under continuous irradiation with microwave field given by Eq.~(\ref{eqn:CPMW}). We trace time-evolutions of the magnetizations for a sufficiently long duration up to $t$=48000 at most. Here we take the microwave amplitude $H^\omega=0.01$ for the simulations. Note that when the microwave field is absent (i.e., $H^\omega$=0), the system exhibits ground-state phases shown in Fig.~\ref{Fig01}(a) where the three magnetic phases with different skyrmion numbers $|N_{\rm sk}|$=2, $|N_{\rm sk}|$=1, and $N_{\rm sk}$=0 successively emerge as $H_z$ increases. We select a ground-state magnetic configuration as an initial state for the time-evolution simulations at a given field strength of $H_z$. We present this ground-state phase diagram also in Fig.~\ref{Fig05} for a reference. We also note that in the static limit of $\omega$=0, all these phases turn into the nontopological phase with $N_{\rm sk}$=0 in the presence of static in-plane magnetic field of $H^\omega$=0.01. Thus, the phases on the $H_z$-axis are all assigned to $N_{\rm sk}$=0.

To discuss the phase diagram in Fig.~\ref{Fig05}, we should first note that a circularly polarized microwave field generates an effective static magnetic field $\pm \omega \bm e_z$ perpendicular to the polarization plane~\cite{Miyake2020,Mochizuki2018}. The amplitude is equal to $\omega$ (i.e., $\hbar\omega/g\mu_{\rm B}$ in real units), while the sign is determined by the sense of the circular polarization, i.e., positive (negative) for the clockwise (counterclockwise) polarization. In the present case, the sign is negative because the microwave field circulating in counterclockwise sense is applied. Thereby, a static component of the total magnetic field acting on the system is $H_z^{\rm tot} = H_z-\omega$. This means that the application of this microwave field effectively work to shift the system towards a low-field regime in the equilibrium phase diagram. Thus, the nonequilibrium phase with $N_{\rm sk}$=0 in the low-frequency regime tends to change into the skyrmion-lattice phase with $|N_{\rm sk}|$=1 and further into the skyrmion-lattice phase with $|N_{\rm sk}|$=2 as $\omega$ increases. Indeed, the skyrmion-lattice phase with $|N_{\rm sk}|$=2 appears in the right area of the phase diagram where $\omega$ is large, whereas the skyrmion-lattice phase with $|N_{\rm sk}|$=1 can emerge when $\omega$ is intermediate in the areas referred to as ``the probabilistic irreversible switching" regime and ``the temporally random fluctuation" regime in the phase diagram.

Importantly, the ways of the emergence of the $|N_{\rm sk}|$=1 phase are distinct from those of the $|N_{\rm sk}|$=2 phase and the $N_{\rm sk}$=0 phase. The latter two phases emerge in a deterministic and irreversible way under irradiation with microwave field. In contrast, the $|N_{\rm sk}|$=1 phase emerges in a probabilistic but irreversible way when $H_z$ is low, whereas either the $|N_{\rm sk}|$=1 phase or the $|N_{\rm sk}|$=2 phase randomly emerges in a temporally fluctuating manner when $H_z$ is high. These peculiar behaviors might be attributable to the energy landscapes characterized by the energy minima and the energy barriers. We expect an energy landscape in Fig.~\ref{Fig05}(b) for the areas where the system enters the $|N_{\rm sk}|$=2 phase or the $N_{\rm sk}$=0 phase in a deterministic and irreversible manner. On the other hand, we expect energy landscapes in Figs.~\ref{Fig05}(c) and (d) for the probabilistic-irreversible-switching regime and the temporally-random-fluctuation regime, respectively, in which the energy minima are nearly degenerate. For the probabilistic irreversible regime, the energy barrier is rather high [see Fig.~\ref{Fig05}(c)], with which the system becomes settled down once it falls into one of the minima, resulting in the probabilistic irreversible switching. In contrast, we expect a small energy barrier in the temporally fluctuating regime where the system fluctuates between the two minima under irradiation with the microwave field. 

\begin{figure}[tbh]
\centering
\includegraphics[scale=1.0]{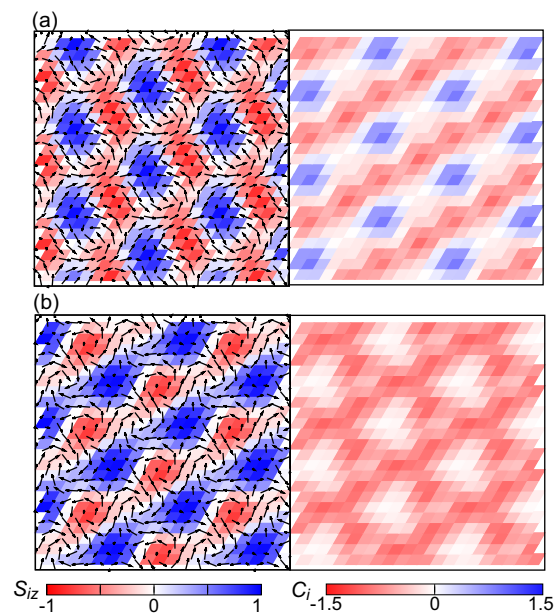}
\caption{(a) Snapshots of the local magnetizations $\bm S_i$ (left panel) and the scalar spin chiralities $C_i$ (right panel) for the dynamical magnetic pattern with a half-integer skyrmion number of $N_{\rm sk}=-1/2$ emerging in the transient process in Fig.~\ref{Fig04}(a). (b) Those for the dynamical magnetic pattern with $N_{\rm sk}=-3/2$ emerging in the transient process in Fig.~\ref{Fig04}(b).}
\label{Fig06}
\end{figure}
Interestingly, we find dynamical formations of topological magnetic patterns with half-integer skyrmion numbers during the switching processes. We present snapshots of their spatial configurations in Figs.~\ref{Fig06}(a) and (b). Specifically, Fig.~\ref{Fig06}(a) shows the dynamical magnetic pattern with $N_{\rm sk}=-1/2$ emerging during the process of magnetic-topology switching from $N_{\rm sk}=-1$ to $N_{\rm sk}=0$ in Fig.~\ref{Fig04}(a), whereas Fig.~\ref{Fig06}(b) shows that with $N_{\rm sk}=-3/2$ emerging during the switching from $N_{\rm sk}=-1$ to $N_{\rm sk}=0$ in Fig.~\ref{Fig04}(b). Topological magnetic textures having a half-integer topological charge are referred to as merons or antimerons, and their crystallized states have been an issue of intensive research~\cite{Yu2018,Hayami2021b,WangZ2021,LinSZ2015,Kharkov2017,Gobel2019}. Clarifications of the observed dynamical topological magnetic patterns in the transient processes are left for future studies.

\section{Discussion}
\begin{figure*}[tbh]
\begin{center}
\includegraphics[scale=1.0]{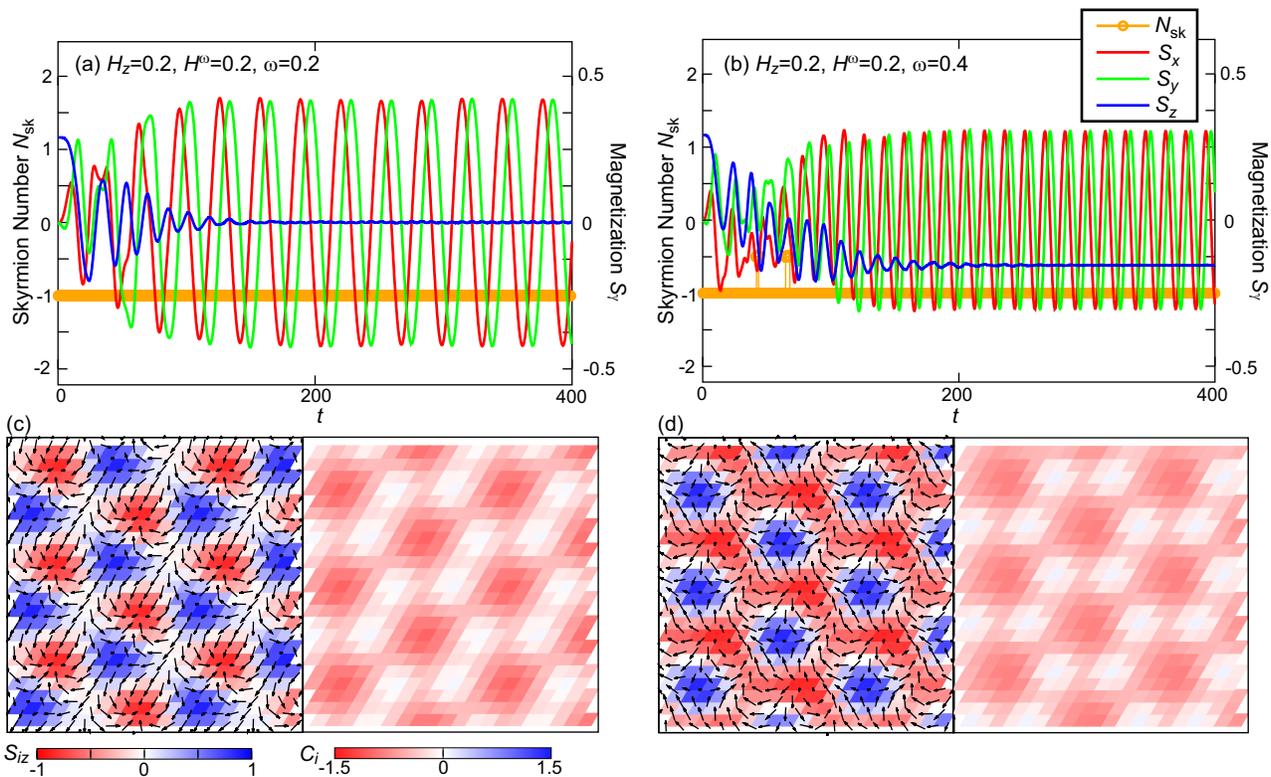}
\caption{Simulated time profiles of the net magnetization $\bm S=(S_x,S_y,S_z)$ and the skrmion number $N_{\rm sk}$ in the effective BBQ model under irradiation with circularly polarized microwave field with (a) $\omega$=0.2 and (b) $\omega$=0.4 when $H_z=0.2$. For both cases, we start with a skyrmion lattice with $N_{\rm sk}=-1$ as an initial state, which is the ground state at $H_z=0.2$. The skyrmion number remains constant to be $N_{\rm sk}=-1$ even after a sufficient duration, indicating that the switching of magnetic topology does not occur in contrast to the case of the Kondo-lattice model. (c),(d) Snapshots of the local magnetizations $\bm S_i$ (left panels) and the scalar spin chiralities $C_i$ (right panels) in the microwave-induced nonequilibrium steady phases at selected moments. The magnetic configuration in (c) corresponds to the bimeron crystal with $N_{\rm sk}=-1$, whereas that in (d) corresponds to the antiskyrmion crystal with $N_{\rm sk}=-1$}
\label{Fig07}
\end{center}
\end{figure*}
In this section, we discuss our results in the light of an effective model introduced in literature~\cite{Ozawa2016,Hayami2017,Akagi2012,Hayami2014b}, which is called the effective bilinear-biquadratic (BBQ) model. This model is derived from the original Kondo-lattice model in Eq.~(\ref{eq:KLM}) using the perturbation expansions when the hopping term dominates the Kondo-coupling term. We argue that effective three-body interactions originating from the third-order perturbation processes might be of essential importance for the observed microwave-induced switching of magnetic topology in the centrosymmetric itinerant magnets. 

The Hamiltonian of the effective BBQ model is given by,
\begin{equation}
\mathcal{H} = \mathcal{H}_{\rm BBQ} + \mathcal{H}_{\rm Zeeman}, 
\end{equation}
with
\begin{equation}
\mathcal{H}_{\rm BBQ} = \sum_{\nu=1}^3 \left[ -J\left|{\bm S}_{{\bm Q}_\nu}\right|^2 + \frac{K}{N}\left|{\bm S}_{{\bm Q}_\nu}\right|^4 \right]. 
\label{eq:BBQ}
\end{equation}
The first term of Eq.~(\ref{eq:BBQ}) represents the effective interactions originating from the second-order perturbation processes, which is often referred to as the RKKY interactions. On the other hand, the second term represents parts of the contributions from the fourth-order perturbation processes~\cite{Akagi2012,Hayami2014b}. Note that contributions from the odd-order perturbation processes usually vanish when magnetic fields are absent because of the time-reversal symmetry, whereas they should appear in the presence of magnetic fields. For the Zeeman-coupling term $\mathcal{H}_{\rm Zeeman}$, we consider both the static perpendicular magnetic field $\bm H_{\rm ex}=(0,0,H_z)$ and the circularly polarized microwave field with amplitude of $H^\omega$. When the microwave field is absent (i.e., $H^\omega=0$), this effective BBQ model is known to host two different skyrmion-lattice phases with $|N_{\rm sk}|=1$ and $|N_{\rm sk}|=2$ as well as the nontopological phase with $N_{\rm sk}=0$ in the ground state depending on the strength of external magnetic field $H_z$.

The coupling constants $J$ and $K$ in Eq.~(\ref{eq:BBQ}) depend on the electronic structures of conduction electrons such as Fermi surfaces and density of states, which are governed by the lattice structure, the transfer integrals, and the electron filling. The values of $J$ and $K$ can be evaluated from the original Kondo-lattice model by the perturbation-expansion calculations in principle. Through the second-order perturbation expansions, we obtain the following formula for the coupling constant $J$,
\begin{equation}
J=J_{\rm K}^2 \sum_{\bm q \in {\rm BZ}} \chi_0(\bm q)e^{i\bm q \cdot \bm r_1}.
\end{equation}
Here $\chi_0(\bm q)$ is the bare susceptibility of conduction electrons, and $\bm r_1$ denotes the Bravis vectors of the triangular lattice. For the Kondo-lattice model in Eq.~(\ref{eq:KLM}) and the parameters used in the present work (i.e., $t_1=-1$, $t_3=0.85$, $J_{\rm K}=-0.5$, and $\mu=-3.5$), we evaluate the value as $J \sim 0.0035 t_1$, which is much smaller than unity and thus supports the validity of the perturbational treatment. We can also evaluate the coupling constant $K$ microscopically from the Kondo-lattice model. However, to get a general insight into the effective BBQ model, we treat the model rather phenomenologically by regarding $J$, $K$, $H_z$ and $H^\omega$ as parameters and by setting the constant $J$ as energy units (i.e., $J=1$) in the following discussion.

Now we examine the microwave-induced magnetization dynamics in this effective BBQ model by deriving a time-evolution equation for the Fourier components of magnetization $\bm S_{\bm q}$. The equation is given by,
\begin{widetext}
\begin{align}
\frac{d\bm S_{\bm q}}{dt}
&=\frac{i}{\hbar}\left[
\mathcal{H}_{\rm BBQ}+\mathcal{H}_{\rm Zeeman}, \bm S_{\bm q}
\right]_{-}
\nonumber \\
&=-\sum_{\nu=1}^3 \left(-J+\frac{K}{N}|\bm S_{\bm Q_\nu}|^2\right)
\left(\bm S_{\bm Q_\nu} \times \bm S_{-\bm Q_\nu+\bm q}
-\bm S_{\bm Q_\nu+\bm q} \times \bm S_{-\bm Q_\nu} \right)
+\bm S_{\bm q} \times \left[\bm H_{\rm ext}+\bm H^\prime(t)\right].
\end{align}
\end{widetext}
By numerically solving the derived equation, we examine two cases with $\omega=0.005$ and $\omega=0.01$, for which we have respectively observed the switching of magnetic topology in the original Kondo-lattice model, i.e., the switching from $N_{\rm sk}=-1$ to $N_{\rm sk}=0$ [Fig.~\ref{Fig04}(a)] and the switching from $N_{\rm sk}=-1$ to $N_{\rm sk}=-2$ [Fig.~\ref{Fig04}(b)]. We set the parameter values as $J=1$, $K=0.5$, $H_z=0.2$ and $H^\omega=0.4$. Surprisingly, the switching of magnetic topology is not observed for both cases [see Figs.~\ref{Fig07}(a) and (b)]. 

The failure of the effective BBQ model in reproduction of the topology switching may be ascribed to ingredients which are incorporated in the original Kondo-lattice model but missing in the effective BBQ model. One missing ingredient is effective interactions originating from the third-order perturbation processes. The effective BBQ model contains only terms bilinear and biquadratic with respect to $\bm S_{\bm Q_\nu}$ and $\bm S_{-\bm Q_\nu}$, which come from the second-order and fourth-order perturbation processes, respectively. We infer that the other-order terms might contribute to the topology switching. Among those terms, the lowest-order terms, i.e., the third-order terms are most likely, which are three-body interactions with respect to $S_{\bm Q_\nu}$ and $S_{-\bm Q_\nu}$ and thus have no time-reversal symmetry. Therefore, the third-order terms are forbidden and should vanish in the absence of magnetic field. However, they are allowed to appear once a magnetic field is applied, although they are not incorporated in the effective BBQ model even under a magnetic field. Contributions of the third-order terms to the magnetization dynamics are described by the following Heisenberg equation of motion for the Fourier component $\bm S_{\bm q}$,
\begin{equation}
\frac{d\bm S_{\bm q}}{dt} 
=\frac{i}{\hbar}\left[ F^{(3)}, \bm S_{\bm q} \right]_.
\end{equation}
There are five kinds of third-order terms $F^{(3)}$ as derived in Ref.~\cite{Hayami2017}, and they turn out to contribute to the time evolution of ${\bm S}_{\bm q}$ as shown in the appendix. Another missing ingredient is contributions from momenta apart from $\bm Q_\nu$. The effective BBQ model contains only the Fourier components of magnetization $\bm S_{\bm Q_\nu}$ with $\bm Q_\nu$ being the modulation wavevectors. However, the bilinear and biquadratic interactions originally have components of other wavevectors. These neglected components of the interactions may play a key role in the topology switching. The issue requires further investigations and is left for future studies.

In the meanwhile, the counterclockwise circularly polarized microwave field considered in the present study generates an effective static magnetic field perpendicular to the polarization plane $-\omega \bm e_z$~\cite{Miyake2020,Mochizuki2018}. Therefore, a perpendicular component of the total magnetic field is $H^{\rm tot}_z=H_z-\omega$. In the case of Fig.~\ref{Fig07}(a), the effective static component $H^{\rm tot}_z$ vanishes (i.e., $H^{\rm tot}_z=0$) because we set $H_z=\omega=0.2$. Under this circumstance, the bimeron crystal with $|N_{\rm sk}|=1$~\cite{Kharkov2017,Gobel2019} appears as a nonequilibrium steady state after a sufficient duration of the microwave irradiation [Fig.~\ref{Fig07}(c)]. Note that the effective BBQ model in Eq.~(\ref{eq:BBQ}) exhibits the skyrmion-lattice phase with $|N_{\rm sk}|=1$ in the equilibrium case when both the static and microwave magnetic fields are absent (i.e., $H_z$=$H^\omega$=0). On the other hand, the effective static component $H^{\rm tot}_z$ is negative (i.e., $H^{\rm tot}_z=-0.2$) because we set $H_z=0.2$ and $\omega=0.4$. We observe the antiskyrmion lattice with $|N_{\rm sk}|=1$ [Fig.~\ref{Fig07}(d)] after a sufficient duration of the microwave irradiation, in which the core magnetizations point upwards.

\section{Conclusion}
In summary, we have theoretically proposed possible microwave-induced switching of magnetic topology in centrosymmetric itinerant magnets by numerically analyzing the magnetization dynamics in the Kondo-lattice model on a triangular lattice using a combined method of the micromagnetic simulation and the kernel polynomial expansion technique. We have demonstrated that the intense excitation of spin-wave mode with circularly polarized microwave field can switch the skyrmion lattice with $|N_{\rm sk}|$=1 into that with $|N_{\rm sk}|$=2 or the nontopological magnetic state with $N_{\rm sk}$=0 depending on the microwave frequency. During these switching processes, transient topological magnetic patterns with half-integer skyrmion numbers of $|N_{\rm sk}|$=1/2 and $|N_{\rm sk}|$=3/2 were observed. These fractionalizations of magnetic topological charges in the dynamical regime are an issue of interest, which should be clarified in future studies. We have found several different switching behaviors under continuous microwave irradiation, that is, deterministic irreversible switching, probabilistic irreversible switching, and temporally random fluctuation depending on the microwave frequency and the strength of external magnetic field, which is attributable to difference of the energy landscape in the dynamical regime. We have also examined the effective BBQ model derived from the perturbation expansions of the Kondo-lattice model and have found that this model fails to reproduce the dynamical switching of magnetic topology. The failure of the effective model containing only the even-order terms conversely indicates that contributions from the odd-order perturbation processes, which break the time-reversal symmetry, are important to understand the observed magnetic topology switching. Here we emphasize that various unfrozen degrees of freedom inherent in centrosymmetric magnets are sources of rich magnetic textures and their controllability with external parameters. Our work will open a new research field to manipulate magnetic topologies in centrosymmetric magnets.

\section{Appendix}
When the electron hoppings dominate the Kondo exchange coupling in the Kondo-lattice model, we can derive effective interactions among the local magnetizations from this model by using the perturbation expansion technique. In the main text, we have argued that the effective interactions originating from the third-order perturbation processes can contribute to the time-evolutions of the Fourier components $S_{\bm q}$. In Ref.~\cite{Hayami2017}, the effective interactions from the third-order perturbation processes have been derived, which have turned out to be three-body interactions among the Fourier components ${\bm S}_{\bm Q_\nu}$ and ${\bm S}_{-\bm Q_\nu}$ with three different ${\bm Q}_\nu$ vectors. The third-order contributions are given by,
\begin{align}
 F^{(3)} &= F^{(3)}_1 + F^{(3)}_2 + F^{(3)}_3 + F^{(3)}_4 + F^{(3)}_5,
\end{align}
with
\begin{widetext}
\begin{align}
F^{(3)}_1 &= -2\frac{J_{\rm K}^3}{\sqrt{N}} \sum_\nu (C_1-C_2) \left[ S_{{\bm Q}_\nu}^z \left( S_{\bm 0}^x S_{-{\bm Q}_\nu}^x + S_{\bm 0}^y S_{-{\bm Q}_\nu}^y \right) + {\rm h.c.} \right], \\
F^{(3)}_2 &= -2\frac{J_{\rm K}^3}{\sqrt{N}} \sum_\nu (C_3-C_4) \left[ S_{{\bm 0}}^z \left( S_{{\bm Q}_\nu}^x S_{-{\bm Q}_\nu}^x + S_{{\bm Q}_\nu}^y S_{-{\bm Q}_\nu}^y \right) + {\rm h.c.} \right], \\
F^{(3)}_3 &= -2\frac{J_{\rm K}^3}{\sqrt{N}} \sum_\nu (C_5-C_6) S_{{\bm 0}}^z S_{{\bm Q}_\nu}^z S_{-{\bm Q}_\nu}^z, \\
F^{(3)}_4 &= -2\frac{J_{\rm K}^3}{\sqrt{N}} (D_1-D_2) \left[ S_{{\bm Q}_1}^z \left( S_{{\bm Q}_2}^x S_{{\bm Q}_3}^x + S_{{\bm Q}_2}^y S_{{\bm Q}_3}^y \right) + {\rm h.c.} \right] \notag \\
& \ \ \ \ \ + ({\bm Q}_1 \rightarrow {\bm Q}_2, \ {\bm Q}_2 \rightarrow {\bm Q}_3, \ {\bm Q}_3 \rightarrow {\bm Q}_1) \notag \\
& \ \ \ \ \ + ({\bm Q}_1 \rightarrow {\bm Q}_3, \ {\bm Q}_2 \rightarrow {\bm Q}_1, \ {\bm Q}_3 \rightarrow {\bm Q}_2), \\
F^{(3)}_5 &= -2\frac{J_{\rm K}^3}{\sqrt{N}} (D_3-D_4) \left[ S_{{\bm Q}_1}^z S_{{\bm Q}_2}^z S_{-{\bm Q}_3}^z + {\rm h.c.} \right].
\end{align}
Here the coefficients $C_\nu$ and $D_\nu$ ($\nu=1,2,3,4$) are calculated by the convolution of Green's functions. Detailed formulas of these coefficients are given in Ref.~\cite{Hayami2017}. 

Equations of the time evolution of ${\bm S}_{\bm q}$ due to the third-order contributions are given by,
\begin{align}
\frac{d{\bm S}_{\bm q}}{dt} &= \frac{i}{\hbar}\left[ F_1^{(3)},{\bm S}_{\bm q} \right]_- \notag \\
    &= 2\frac{J_{\rm K}^3}{N}\sum_\nu(C_1-C_2) \left[ \left({\bm S}_{{\bm q}+{\bm Q}_\nu} \times {\bm e}_z\right)\left( S^x_{\bf 0}S^x_{-{\bm Q}_\nu} + S^y_{\bf 0}S^y_{-{\bm Q}_\nu}\right) \right. \notag \\
    & \ \ \ \ \ \left. + \left(
    \begin{array}{c}
    -S_{{\bm Q}_\nu}^z\left( S^y_{\bf 0}S^z_{{\bm q}-{\bm Q}_\nu} + S^z_{{\bm q}}S^y_{-{\bm Q}_\nu} \right) \\
    \ \ S_{{\bm Q}_\nu}^z\left( S^x_{\bf 0}S^z_{{\bm q}-{\bm Q}_\nu} + S^z_{{\bm q}}S^x_{-{\bm Q}_\nu} \right) \\
    -S_{{\bm Q}_\nu}^z\left( S^x_{\bf 0}S^y_{{\bm q}-{\bm Q}_\nu} + S^y_{{\bm q}}S^x_{-{\bm Q}_\nu} \right) + S_{{\bm Q}_\nu}^z\left( S^y_{\bf 0}S^x_{{\bm q}-{\bm Q}_\nu} + S^x_{{\bm q}}S^y_{-{\bm Q}_\nu} \right)
    \end{array}
    \right) \right] \notag \\
    & \ \ \ \ \ + ({\bm Q}_\nu \rightarrow -{\bm Q}_\nu),
\end{align}
\begin{align}
    \frac{d{\bm S}_{\bm q}}{dt} &= \frac{i}{\hbar}\left[ F_2^{(3)},{\bm S}_{\bm q} \right]_- \notag \\
    &= 2\frac{J_{\rm K}^3}{N}\sum_\nu(C_3-C_4) \left[ \left({\bm S}_{{\bm q}} \times {\bm e}_z\right)\left( S^x_{{\bm Q}_\nu}S^x_{-{\bm Q}_\nu} + S^y_{{\bm Q}_\nu}S^y_{-{\bm Q}_\nu}\right) \right. \notag \\
    & \ \ \ \ \ \left. + \left(
    \begin{array}{c}
    -S_{\bf 0}^z\left( S^y_{{\bm Q}_\nu}S^z_{{\bm q}-{\bm Q}_\nu} + S^z_{{\bm q}+{\bm Q}_\nu}S^y_{-{\bm Q}_\nu} \right) \\
    \ \ S_{\bf 0}^z\left( S^x_{{\bm Q}_\nu}S^z_{{\bm q}-{\bm Q}_\nu} + S^z_{{\bm q}+{\bm Q}_\nu}S^x_{-{\bm Q}_\nu} \right) \\
    -S_{\bf 0}^z\left( S^x_{{\bm Q}_\nu}S^y_{{\bm q}-{\bm Q}_\nu} + S^y_{{\bm q}+{\bm Q}_\nu}S^x_{-{\bm Q}_\nu} \right) + S_{\bf 0}^z\left( S^y_{{\bm Q}_\nu}S^x_{{\bm q}-{\bm Q}_\nu} + S^x_{{\bm q}+{\bm Q}_\nu}S^y_{-{\bm Q}_\nu} \right)
    \end{array}
    \right) \right] \notag \\
    & \ \ \ \ \ + ({\bm Q}_\nu \rightarrow -{\bm Q}_\nu),
\end{align}
\begin{align}
    \frac{d{\bm S}_{\bm q}}{dt} &= \frac{i}{\hbar}\left[ F_3^{(3)},{\bm S}_{\bm q} \right]_- \notag \\
    &= 2\frac{J_{\rm K}^3}{N}\sum_\nu(C_5-C_6) \left(
    \begin{array}{c}
    \ \ {\bm S}^x_{{\bm q}}{\bm S}^z_{{\bm Q}_\nu}{\bm S}^z_{-{\bm Q}_\nu} + {\bm S}^z_{\bf 0}{\bm S}^x_{{\bm q}+{\bm Q}_\nu}{\bm S}^z_{-{\bm Q}_\nu} + {\bm S}^z_{\bf 0}{\bm S}^z_{{\bm Q}_\nu}{\bm S}^x_{{\bm q}-{\bm Q}_\nu} \\
    -{\bm S}^y_{{\bm q}}{\bm S}^z_{{\bm Q}_\nu}{\bm S}^z_{-{\bm Q}_\nu} - {\bm S}^z_{\bf 0}{\bm S}^y_{{\bm q}+{\bm Q}_\nu}{\bm S}^z_{-{\bm Q}_\nu} - {\bm S}^z_{\bf 0}{\bm S}^z_{{\bm Q}_\nu}{\bm S}^y_{{\bm q}-{\bm Q}_\nu}  \\
    0
    \end{array}
    \right),
\end{align}
\begin{align}
\frac{d{\bm S}_{\bm q}}{dt} &= \frac{i}{\hbar}\left[ F_4^{(3)},{\bm S}_{\bm q} \right]_- \notag \\
    &= 2\frac{J_{\rm K}^3}{N}(D_1-D_2) \left[ \left({\bm S}_{{\bm q}+{\bm Q}_1} \times {\bm e}_z\right)\left( S^x_{{\bm Q}_2}S^x_{{\bm Q}_3} + S^y_{{\bm Q}_2}S^y_{{\bm Q}_3}\right) \right. \notag \\
    & \ \ \ \ \ \left. + \left(
    \begin{array}{c}
    -S_{{\bm Q}_1}^z\left( S^y_{{\bm Q}_2}S^z_{{\bm q}+{\bm Q}_3} + S^z_{{\bm q}+{\bm Q}_2}S^y_{{\bm Q}_3} \right) \\
    \ \ S_{{\bm Q}_1}^z\left( S^x_{{\bm Q}_2}S^z_{{\bm q}+{\bm Q}_3} + S^z_{{\bm q}+{\bm Q}_2}S^x_{{\bm Q}_3} \right) \\
    -S_{{\bm Q}_1}^z\left( S^x_{{\bm Q}_2}S^y_{{\bm q}+{\bm Q}_3} + S^y_{{\bm q}+{\bm Q}_2}S^x_{{\bm Q}_3} \right) + S_{{\bm Q}_1}^z\left( S^y_{{\bm Q}_2}S^x_{{\bm q}+{\bm Q}_3} + S^x_{{\bm q}+{\bm Q}_2}S^y_{{\bm Q}_3} \right)
    \end{array}
    \right) \right] \notag \\
    & \ \ \ \ \ + ({\bm Q}_1 \rightarrow {\bm Q}_2, \ {\bm Q}_2 \rightarrow {\bm Q}_3, \ {\bm Q}_3 \rightarrow {\bm Q}_1) \notag \\
    & \ \ \ \ \ + ({\bm Q}_1 \rightarrow {\bm Q}_3, \ {\bm Q}_2 \rightarrow {\bm Q}_1, \ {\bm Q}_3 \rightarrow {\bm Q}_2) \notag \\
    & \ \ \ \ \ + ({\bm Q}_1 \rightarrow -{\bm Q}_1, \ {\bm Q}_2 \rightarrow -{\bm Q}_2, \ {\bm Q}_3 \rightarrow -{\bm Q}_3) \notag \\
    & \ \ \ \ \ + ({\bm Q}_1 \rightarrow -{\bm Q}_2, \ {\bm Q}_2 \rightarrow -{\bm Q}_3, \ {\bm Q}_3 \rightarrow -{\bm Q}_1) \notag \\
    & \ \ \ \ \ + ({\bm Q}_1 \rightarrow -{\bm Q}_3, \ {\bm Q}_2 \rightarrow -{\bm Q}_1, \ {\bm Q}_3 \rightarrow -{\bm Q}_2),
\end{align}
\begin{align}
\frac{d{\bm S}_{\bm q}}{dt} &= \frac{i}{\hbar}\left[ F_5^{(3)},{\bm S}_{\bm q} \right]_- \notag \\
    &= 2\frac{J_{\rm K}^3}{N}(D_3-D_4) \left(
    \begin{array}{c}
    \ \ {\bm S}^x_{{\bm q}+{\bm Q}_1}{\bm S}^z_{{\bm Q}_2}{\bm S}^z_{-{\bm Q}_3} + {\bm S}^z_{{\bm Q}_1}{\bm S}^x_{{\bm q}+{\bm Q}_2}{\bm S}^z_{-{\bm Q}_3} + {\bm S}^z_{{\bm Q}_1}{\bm S}^z_{{\bm Q}_2}{\bm S}^x_{{\bm q}-{\bm Q}_3} \\
    -{\bm S}^y_{{\bm q}+{\bm Q}_1}{\bm S}^z_{{\bm Q}_2}{\bm S}^z_{-{\bm Q}_3} - {\bm S}^z_{{\bm Q}_1}{\bm S}^y_{{\bm q}+{\bm Q}_2}{\bm S}^z_{-{\bm Q}_3} - {\bm S}^z_{{\bm Q}_1}{\bm S}^z_{{\bm Q}_2}{\bm S}^y_{{\bm q}-{\bm Q}_3} \\
    0
    \end{array}
    \right) \notag \\
    & \ \ \ \ \ + ({\bm Q}_1 \rightarrow {\bm Q}_2, \ {\bm Q}_2 \rightarrow {\bm Q}_3, \ {\bm Q}_3 \rightarrow {\bm Q}_1) \notag \\
    & \ \ \ \ \ + ({\bm Q}_1 \rightarrow {\bm Q}_3, \ {\bm Q}_2 \rightarrow {\bm Q}_1, \ {\bm Q}_3 \rightarrow {\bm Q}_2) \notag \\
    & \ \ \ \ \ + ({\bm Q}_1 \rightarrow -{\bm Q}_1, \ {\bm Q}_2 \rightarrow -{\bm Q}_2, \ {\bm Q}_3 \rightarrow -{\bm Q}_3) \notag \\
    & \ \ \ \ \ + ({\bm Q}_1 \rightarrow -{\bm Q}_2, \ {\bm Q}_2 \rightarrow -{\bm Q}_3, \ {\bm Q}_3 \rightarrow -{\bm Q}_1) \notag \\
    & \ \ \ \ \ + ({\bm Q}_1 \rightarrow -{\bm Q}_3, \ {\bm Q}_2 \rightarrow -{\bm Q}_1, \ {\bm Q}_3 \rightarrow -{\bm Q}_2).
\end{align}
\end{widetext}

\section{Acknowledgement}
This work is supported by Japan Society for the Promotion of Science KAKENHI (Grant No. 16H06345, No. 19H00864, No. 19K21858, and No. 20H00337), CREST, the Japan Science and Technology Agency (Grant No. JPMJCR20T1), a Research Grant in the Natural Sciences from the Mitsubishi Foundation, and a Waseda University Grant for Special Research Projects (Project No. 2020C-269 and No. 2021C-566). A part of the numerical simulations was performed at the Supercomputer Center of the Institute for Solid State Physics in the University of Tokyo.


\begin{thebibliography}{999}
\bibitem{Muhlbauer2009} S. M{\"u}hlbauer, B. Binz, F. Jonietz, C. Pfleiderer, A. Rosch,  A. Neubauer, R. Georgii, and P. B{\"o}ni, Science {\bf 323}, 915 (2009).

\bibitem{Yu2010} X. Z. Yu, Y. Onose, N. Kanazawa, J. H. Park, J. H. Han, Y. Matsui, N. Nagaosa, and Y. Tokura, Nature {\bf 465}, 901 (2010).

\bibitem{Heinze2011}S. Heinze, K. von Bergmann, M. Menzel, J. Brede, A. Kubetzka,
R. Wiesendanger, G. Bihlmayer, and S. Blugel, Nat. Phys. 7, 713 (2011).

\bibitem{Kezsmarki2015}I. K\'ezsmarki, S. Bordacs, P. Milde, E. Neuber, L. M. Eng, J. S. White, H. M. Ronnow, C. D. Dewhurst, M. Mochizuki, K. Yanai, H. Nakamura, D. Ehlers, V. Tsurkan, and A. Loidl, Nat. Mater. {\bf 14}, 1116 (2015).

\bibitem{Yu2014}X. Z. Yu, Y. Tokunaga, Y. Kaneko, W. Z. Zhang, K. Kimoto, Y. Matsui, Y. Taguchi, and Y. Tokura, Nat. Commun. {\bf 5}, 3198 (2014).

\bibitem{Nayak2017}A. K. Nayak, V. Kumar, T. Ma, P. Werner, E. Pippel, R. Sahoo, F. Damay, U. K. R\"osler, C. Felser, and S. S. Parkin, Nature {\bf 548}, 561 (2017).

\bibitem{Yu2018}X. Z. Yu, W. Koshibae, Y. Tokunaga, K. Shibata, Y. Taguchi, N. Nagaosa, and Y. Tokura, Nature {\bf 564}, 95 (2018).

\bibitem{Milde2013}P. Milde, D. K\"ohler, J. Seidel, L. M. Eng, A. Bauer, A. Chacon, J. Kindervater, S. M\"uhlbauer, C. Pfleiderer, S. Buhrandt, C. Sch\"utte, and A. Rosch, Science {\bf 340}, 1076 (2013).

\bibitem{Schutte2014}C. Sch\"utte and A. Rosch, Phys. Rev. B {\bf 90}, 174432 (2014).

\bibitem{Kanazawa2012}N. Kanazawa, J.-H. Kim, D. S. Inosov, J. S. White, N. Egetenmeyer, J. L. Gavilano, S. Ishiwata, Y. Onose, T. Arima, B. Keimer, and Y. Tokura, Phys. Rev. B {\bf 86}, 134425 (2012).

\bibitem{Fujishiro2019}Y. Fujishiro, N. Kanazawa, T. Nakajima, X. Z. Yu, K. Ohishi, Y. Kawamura, K. Kakurai, T. Arima, H. Mitamura, A. Miyake, K. Akiba, M. Tokunaga, A. Matsuo, K. Kindo, T. Koretsune, R. Arita, and Y. Tokura, Nat. Commun. {\bf 10}, 1059 (2019).

\bibitem{Ishiwata2020} S. Ishiwata, T. Nakajima, J.-H. Kim, D. S. Inosov, N. Kanazawa, J. S. White, J. L. Gavilano, R. Georgii, K. M. Seemann, G. Brandl, P. Manuel, D. D. Khalyavin, S. Seki, Y. Tokunaga, M. Kinoshita, Y. W. Long, Y. Kaneko, Y. Taguchi, T. Arima, B. Keimer, and Y. Tokura, Phys. Rev. B {\bf 101}, 134406 (2020).

\bibitem{LiuY2018}Y. Liu, R. K. Lake, J. Zang, Phys. Rev. B {\bf 98}, 174437 (2018).
\bibitem{Seki2016}S. Seki and M. Mochizuki, {\it Skyrmions in Magnetic Materials} (Springer, Berlin, 2016).

\bibitem{Nagaosa2013}N. Nagaosa and Y. Tokura, Nat. Nanotech. {\bf 8}, 899 (2013).

\bibitem{Gobel2021}B. G\"{o}bel, I. Mertig, and O. A. Tretiakov, Phys. Rep. {\bf 895}, 1 (2021).

\bibitem{Fert2013}A. Fert, V. Cros, and J. Sampaio, Nat. Nanotech. {\bf 8}, 152 (2013).

\bibitem{Finocchio2016}G. Finocchio, F. Buttner, R. Tomasello, M. Carpentieri, and M. Kl\"aui, J. Phys. D: Appl. Phys. {\bf 49}, 423001 (2016).

\bibitem{Fert2017}A. Fert, N. Reyren, and V. Cros, Nat. Rev. Mat. {\bf 2}, 17031 (2017).

\bibitem{Everschor2019}K. Everschor-Sitte, J. Masell, R. M. Reeve, and M. Kl\"aui, J. Appl. Phys. {\bf 124}, 240901 (2018).

\bibitem{Tokura2021}Y. Tokura and N. Kanazawa, Chem. Rev. {\bf 121}, 2857 (2021).
\bibitem{Dzyaloshinsky1958}I. Dzyaloshinsky, J. Phys. Chem. Solids {\bf 4}, 241 (1958).

\bibitem{Moriya1960a}T. Moriya, Phys. Rev. {\bf 120}, 91 (1960).

\bibitem{Moriya1960b}T. Moriya, Phys. Rev. Lett. {\bf 4}, 228 (1960).
\bibitem{Bogdanov1989}A. N. Bogdanov and D. A. Yablonskii, Sov. Phys. JETP {\bf 68}, 101 (1989).

\bibitem{Bogdanov1994}A. Bogdanov and A. Hubert, J. Magn. Magn. Mater. {\bf 138}, 255 (1994).

\bibitem{Tanaka2020}K. Tanaka, R. Sugawara, and M. Mochizuki, Phys. Rev. Mat. {\bf 4}, 034404 (2020).
\bibitem{Hayami2021R}S. Hayami and Y. Motome, arXiv:2103.10647.

\bibitem{Hayami2014a}S. Hayami, T. Misawa, Y. Yamaji, and Y. Motome, Phys. Rev. B {\bf 89}, 085124 (2014).

\bibitem{Ozawa2016}R. Ozawa, S. Hayami, K. Barros, G.-W. Chern, Y. Motome, and C. D. Batista, J. Phys. Soc. Jpn. {\bf 85} 103703 (2016).

\bibitem{Ozawa2017a}R. Ozawa, S. Hayami, and Y. Motome, Phys. Rev. Lett. {\bf 118}, 147205 (2017).

\bibitem{Hayami2017}S. Hayami, R. Ozawa, and Y. Motome, Phys. Rev. B {\bf 95}, 224424 (2017).

\bibitem{Gobel2017}B. G\"{o}bel, A. Mook, J. Henk, and I. Mertig, Phys. Rev. B {\bf 96}, 060406(R) (2017).

\bibitem{Wang2020}Z. Wang, Y. Su, S.-Z. Lin, and C. D. Batista, Phys. Rev. Lett. {\bf 124}, 207201 (2020).

\bibitem{Hayami2020a}S. Hayami and R. Yambe, J. Phys. Soc. Jpn. {\bf 89}, 103702 (2020).

\bibitem{Hayami2021a}S. Hayami and Y. Motome, Phys. Rev. B  {\bf 103}, 024439 (2021).

\bibitem{Hayami2021b}S. Hayami and Y. Motome, Phys. Rev. B {\bf 103}, 054422 (2021).
\bibitem{Nakazawa2019}K. Nakazawa and H. Kohno, Phys. Rev. B {\bf 99}, 174425 (2019).

\bibitem{Hayami2020b}S. Hayami, T. Okubo, and Y. Motome, arXiv:2005.03168.
\bibitem{Hayami2018}S. Hayami and Y. Motome, Phys. Rev. Lett. {\bf 121}, 137202 (2018).

\bibitem{Okada2018}K. N. Okada, Y. Kato, and Y. Motome, Phys. Rev. B {\bf 98}, 224406 (2018).

\bibitem{Okumura2020} S. Okumura, S. Hayami, Y. Kato, and Y. Motome, Phys. Rev. B {\bf 101}, 144416 (2020).

\bibitem{Shimizu2021a}K. Shimizu, S. Okumura, Y. Kato, and Y. Motome, Phys. Rev. B {\bf 103}, 054427 (2021).
\bibitem{Ruderman1954}M. A. Ruderman and C. Kittel, Phys. Rev. {\bf 96}, 99 (1954).

\bibitem{Kasuya1956}T. Kasuya, Prog. Theor. Phys. {\bf 16}, 45 (1956).

\bibitem{Yosida1957}K. Yosida, Phys. Rev. {\bf 106}, 893 (1957).
\bibitem{Shimizu2021b}K. Shimizu, S. Okumura, Y. Kato, and Y. Motome, Phys. Rev. B {\bf 103}, 184421 (2021).

\bibitem{Kurumaji2019}T. Kurumaji, T. Nakajima, M. Hirschberger, A. Kikkawa, Y. Yamasaki, H. Sagayama, H. Nakao, Y. Taguchi, T. Arima, and Y. Tokura, Science {\bf 365}, 914 (2019).

\bibitem{Hirschberger2020a}M. Hirschberger, L. Spitz, T. Nomoto, T. Kurumaji, S. Gao, J. Masell, T. Nakajima, A. Kikkawa, Y. Yamasaki, H. Sagayama, H. Nakao, Y. Taguchi, R. Arita, T.-h. Arima, and Y. Tokura, Phys. Rev. Lett. {\bf 125}, 076602 (2020).

\bibitem{Hirschberger2020b}M. Hirschberger, T. Nakajima, M. Kriener, T. Kurumaji, L. Spitz, S. Gao, A. Kikkawa, Y. Yamasaki, H. Sagayama, H. Nakao, S. Ohira-Kawamura, Y. Taguchi, T.-h. Arima, and Y. Tokura, Phys. Rev. B {\bf 101}, 220401(R) (2020).

\bibitem{Nomoto2020}T. Nomoto, T. Koretsune, and R. Arita, Phys. Rev. Lett. {\bf 125}, 117204 (2020).

\bibitem{Hirschberger2019}M. Hirschberger, T. Nakajima, S. Gao, L. Peng, A. Kikkawa, T. Kurumaji, M. Kriener, Y. Yamasaki, H. Sagayama, H. Nakao, K. Ohishi, K. Kakurai, Y. Taguchi, X. Yu, T.-h. Arima, and Y. Tokura, Nat. Commun. {\bf 10}, 5831 (2019).

\bibitem{Hirschberger2021}M. Hirschberger, S. Hayami, and Y. Tokura, New J. Phys. {\bf 23} 023039 (2021).

\bibitem{Khanh2020}N. D. Khanh, T. Nakajima, X. Yu, S. Gao, K. Shibata, M. Hirschberger, Y. Yamasaki, H. Sagayama, H. Nakao, L. Peng, K. Nakajima, R. Takagi, T. Arima, Y. Tokura, and S. Seki, Nat. Nanotech. {\bf 15}, 444 (2020).

\bibitem{Yasui2020}Y. Yasui, C. J. Butler, N. D. Khanh, S. Hayami, T. Nomoto, T. Hanaguri, Y. Motome, R. Arita, T.-H. Arima, Y. Tokura, and S. Seki, Nat. Commun. {\bf 11}, 5925 (2020).

\bibitem{SZLin2013}S.-Z. Lin, C. Reichhardt, C. D. Batista, and A. Saxena, Phys. Rev. Lett. {\bf 113}, 207202 (2013).
\bibitem{Mochizuki2012}M. Mochizuki, Phys. Rev. Lett. {\bf 108}, 017601 (2012).

\bibitem{Schwarze2015}T. Schwarze, J. Waizner, M. Garst, A. Bauer, I. Stasinopoulos, H. Berger, C. Pfleiderer, and D. Grundler, Nat. Mater. {\bf 14}, 478 (2015).

\bibitem{LinSZ2014}S.-Z. Lin, C. D. Batista, and A. Saxena, Phys. Rev. B {\bf 89}, 024415 (2014).

\bibitem{McKeever2019}B. F. McKeever, D. R. Rodrigues, D. Pinna, Ar. Abanov, J. Sinova, and K. Everschor-Sitte, Phys. Rev. B {\bf 99}, 054430 (2019).

\bibitem{LinSZ2019}S.-Z. Lin, J.-X. Zhu, and A. Saxena, Phys. Rev. B {\bf 99}, 140408(R) (2019).

\bibitem{Seki2020}S. Seki, M. Garst, J. Waizner, R. Takagi, N. D. Khanh, Y. Okamura, K. Kondou, F. Kagawa, Y. Otani, and Y. Tokura, Nat. Commun. {\bf 11}, 256 (2020).
\bibitem{Mochizuki2015}M. Mochizuki and S. Seki, J. Phys.: Condens. Matter {\bf 27}, 503001 (2015).

\bibitem{Garst2017}M. Garst, J. Waizner, and D. Grundler, J. Phys. D: Appl. Phys. {\bf 50}, 293002 (2017).

\bibitem{Lonsky2020}M. Lonsky and A. Hoffmann, APL Materials, {\bf 8}, 100903 (2020).
\bibitem{Mochizuki2013}M. Mochizuki and S. Seki, Phys. Rev. B {\bf 87}, 134403 (2013).

\bibitem{Okamura2013}Y. Okamura, F. Kagawa, M. Mochizuki, M. Kubota, S. Seki, S. Ishiwata, M. Kawasaki, Y. Onose, and Y. Tokura, Nat. Commun. {\bf 4}, 2391 (2013).

\bibitem{MaF2015}F. Ma, Y. Zhou, H. B. Braun, and W. S. Lew, Nano Lett. {\bf 15}, 4029 (2015).

\bibitem{WangK2020}X. Wang, Y. Nie, Q. Xia, and G. Guo, J. Appl. Phys. {\bf 128}, 063901 (2020).

\bibitem{WangW2015}W. Wang, M. Beg, B. Zhang, W. Kuch, and H. Fangohr, Phys. Rev. B {\bf 92}, 020403(R) (2015).

\bibitem{Ikka2018}M. Ikka, A. Takeuchi, and M. Mochizuki, Phys. Rev. B {\bf 98}, 184428 (2018).

\bibitem{Takeuchi2018}A. Takeuchi, and M. Mochizuki, Appl. Phys. Lett. {\bf 113}, 072404 (2018).

\bibitem{SongC2019}C. Song, C. Jin, Y. Ma, J. Wang, H. Xia, J. Wang, and Q. Liu, J. Phys. D: Appl. Phys. {\bf 52}, 435001 (2019).

\bibitem{YuanHY2019}H. Y. Yuan, X. S. Wang, M.-H. Yung, and X. R. Wang, Phys. Rev. B {\bf 99}, 014428 (2019).

\bibitem{Koide2019}T. Koide, A. Takeuchi, and M. Mochizuki, Phys. Rev. B {\bf 100}, 014408 (2019).

\bibitem{Takeuchi2019}A. Takeuchi, S. Mizushima, and M. Mochizuki, Sci. Rep. {\bf 9}, 9528 (2019).

\bibitem{Miyake2020}M. Miyake and M. Mochizuki, Phys. Rev. B {\bf 101}, 094419 (2020).

\bibitem{XingX2020}X. Xing, Y. Zhou, and H.B. Braun, Phys. Rev. Appl. {\bf 13}, 034051 (2020).

\bibitem{Ser2020}N. D. Ser, L. Heinen, and A. Rosch, arXiv:2012.11548 (To appear in SciPost).
\bibitem{Okubo2012}T. Okubo, S. Chung, and H. Kawamura, Phys. Rev. Lett. {\bf 108}, 017206 (2012).

\bibitem{Kamiya2014}Y. Kamiya and C. D. Batista, Phys. Rev. X {\bf 4}, 011023 (2014).

\bibitem{Leonov2015}A. O. Leonov and M. Mostovoy, Nat. Commun. {\bf 6}, 8275 (2015).

\bibitem{LinSZ2016}S.-Z. Lin and S. Hayami, Phys. Rev. B {\bf 93}, 064430 (2016).

\bibitem{Hayami2016a}S. Hayami, S.-Z. Lin, and C. D. Batista, Phys. Rev. B {\bf 93}, 184413 (2016).

\bibitem{Hayami2016b}S. Hayami, S.-Z. Lin, Y. Kamiya, and C. D. Batista, Phys. Rev. B {\bf 94}, 174420 (2016).

\bibitem{LinSZ2018}S.-Z. Lin and C. D. Batista, Phys. Rev. Lett. {\bf 120}, 077202 (2018).

\bibitem{Lohani2019}V. Lohani, C. Hickey, J. Masell, and A. Rosch, Phys. Rev. X {\bf 9}, 041063 (2019).

\bibitem{WangZ2021}Z. Wang, Y. Su, S. Z. Lin, and C. D. Batista, Phys. Rev. B {\bf 103}, 104408 (2021).
\bibitem{Motome1999}Y. Motome and N. Furukawa, J. Phys. Soc. Jpn. {\bf 68}, 3853 (1999).

\bibitem{Weisse2006}A. Wei{\ss}e, G. Wellein, A. Alvermann, and H. Fehske, Rev. Mod. Phys. {\bf 78}, 275 (2006).

\bibitem{Barros2013}K. Barros and Y. Kato, Phys. Rev. B {\bf 88}, 235101 (2013).

\bibitem{Wang2016} Z. Wang, K. Barros, G.-W. Chern, D. L. Maslov, and C. D. Batista, Phys. Rev. Lett. {\bf 117}, 206601 (2016).

\bibitem{Ozawa2017b}R. Ozawa, S. Hayami, K. Barros, and Y. Motome, Phys. Rev. B {\bf 96}, 094417 (2017).

\bibitem{Chern2018}G.-W. Chern, K. Barros, Z. Wang, H. Suwa, and C. D. Batista, Phys. Rev. B {\bf 97}, 035120 (2018).

\bibitem{Wang2018}Z. Wang, G.-W. Chern, C. D. Batista, and K. Barros, J. Chem. Phys. {\bf 148}, 094107 (2018).

\bibitem{Tang2012} J. M. Tang and Y. Saad, Numer. Linear Algebra Appl. {\bf 19}, 485 (2012).
\bibitem{Kim2020}J.-V. Kim, and J. Mulkers, IOP SciNotes {\bf 1}, 025211 (2020).
\bibitem{Mochizuki2018} M. Mochizuki, K. Ihara, J. Ohe, and A. Takeuchi, Appl. Phys. Lett. {\bf 112}, 122401 (2018).
\bibitem{LinSZ2015}S.-Z. Lin, A. Saxena, and C. D. Batista, Phys. Rev. B {\bf 91}, 224407 (2015).
\bibitem{Kharkov2017}Y. A. Kharkov, O. P. Sushkov, and M. Mostovoy, Phys. Rev. Lett. {\bf 119}, 207201 (2017).
\bibitem{Gobel2019}B. G\"{o}bel, A. Mook, J. Henk, I. Mertig, and O. A. Tretiakov, Phys. Rev. B {\bf 99}, 060407(R) (2019).

\bibitem{Akagi2012}Y. Akagi, M. Udagawa, and Y. Motome, Phys. Rev. Lett. {\bf 108}, 096401 (2012).

\bibitem{Hayami2014b}S. Hayami and Y. Motome, Phys. Rev. B {\bf 90}, 060402(R) (2014).
\end{thebibliography}
\end{document}